\documentclass[10pt,conference]{IEEEtran}

\IEEEoverridecommandlockouts
\usepackage{graphicx} 
\usepackage{color}
\usepackage{array}
\usepackage{bm}
\usepackage{amsmath}
\usepackage{amssymb}
\usepackage{amsfonts}
\usepackage{algorithm}
\usepackage{algpseudocode}
\usepackage{graphicx}
\graphicspath{ {./Fig/} }
\usepackage{caption}
\usepackage{comment}
\usepackage{xspace}
\usepackage{algpseudocode}
\usepackage{subfig}
\usepackage{float}
\usepackage{cite}
\usepackage[colorlinks=true,linkcolor=black, citecolor=black, urlcolor=black]{hyperref}
\usepackage{adjustbox}

\newcommand{\sol}{{ZEST}\xspace}
\newcommand{\atool}{{SANE}\xspace}

\title{\sol: Attention-based Zero-Shot Learning for Unseen IoT Device Classification}

\begin{document}

\author{

\IEEEauthorblockN{
Binghui Wu, Philipp Gysel, Dinil Mon Divakaran, and Mohan Gurusamy}

\thanks{Binghui Wu and Mohan Gurusamy (Senior Member, IEEE) are with National University of Singapore (NUS); e-mail-id: e0675868@u.nus.edu, gmohan@ieee.org. Philipp Gysel and Dinil Mon Divakaran (Senior Member, IEEE) are with Acronis Research; e-mail-id: Philipp.Gysel@acronis.com, dinil@comp.nus.edu.sg.}
}

\maketitle

\begin{abstract}
Recent research works have proposed machine learning models for classifying IoT devices connected to a network. However, there is still a practical challenge of not having all devices (and hence their traffic) available during the training of a model. This essentially means, during the operational phase, we need to classify new devices {\it not seen} in the training phase. To address this challenge, we propose \sol---a ZSL (zero-shot learning) framework based on self-attention for classifying both {\it seen} and {\it unseen} devices. \sol consists of i)~a self-attention based network feature extractor, termed \atool, for extracting latent space representations of IoT traffic, ii)~a generative model that trains a decoder using latent features to generate pseudo data, and iii)~a supervised model that is trained on the generated pseudo data for classifying devices. We carry out extensive experiments on real IoT traffic data; our experiments demonstrate i)~\sol achieves significant improvement (in terms of accuracy) over the baselines; ii)~\atool is able to better extract meaningful representations than LSTM which has been commonly used for modeling network traffic.

\end{abstract}


\section{Introduction}

\IEEEPARstart{O}{ffices}, homes, and enterprises in various industry verticals have numerous IoT devices connected to their networks, including smart thermostats, hubs, lighting systems, alarms, TVs, and wearable devices. While IoT devices offer new and efficient services, they also present security threats. Currently, manufacturers do not follow a standard framework to announce the device identities and their functionalities. The lack of standardization often results in vulnerabilities left open for different kinds of attacks~\cite{Understanding-Mira-2017, hajime-NDSS-2019}. An important first step in securing IoT devices is to identify the different types of devices operating in a home/office environment. The challenge to the requirements arises from the constantly evolving landscape of IoT devices and their network behaviors. Traditional static methods struggle to adapt to changing device behaviors, recognize unknown devices, and capture complex communication patterns~\cite{2018-Iot-behavior-print}. Consequently, monitoring network traffic dynamically is now the most practical method for identifying devices and ensuring their security.

IoT fingerprinting is a well-studied problem~\cite{IoT_traffic_Apthorpe, DBLP_Markus, DBLP_Sivanathan, LSTM-device-fingerprinting-2020, unsuper_2022_meditcom, ICC_unsuper,DEFT-2019, Comsnet-2019, r5, cost-aware-classification-IoT-2021, CNN_semi_2020, iPET-PETS-2023, semi_detection_2023}, but there are still open challenges in practical settings. Many existing works take a supervised approach, thereby dealing only with known devices~\cite{IoT_traffic_Apthorpe, DBLP_Markus, DBLP_Sivanathan, LSTM-device-fingerprinting-2020, cost-aware-classification-IoT-2021}. 
With the number of IoT devices expected to grow to tens of billions, new device types will continue to enter the market, making it impractical to assume that traffic of all devices will be available in advance to train a machine learning model. The challenge remains to identify devices not present in the training set, which we refer to as ``{\it unseen}" devices. Conversely, ``{\it seen}" devices refer to those devices that have labeled examples available during the model's training phase. We need a system that can classify {\it unseen} IoT devices, in addition to {\it seen} devices.

\begin{figure}
    \centering
    \subfloat{\includegraphics[scale=0.23]{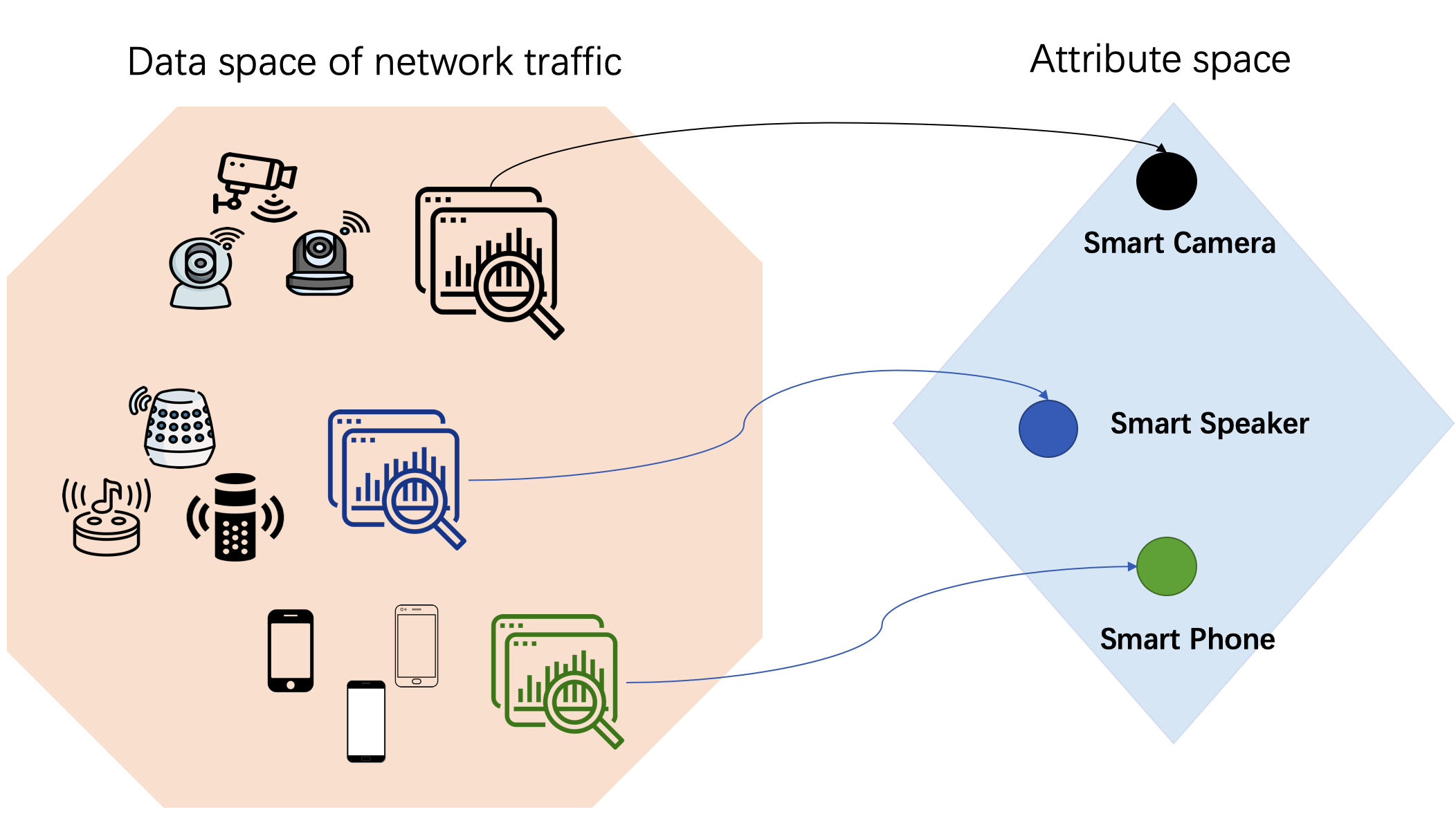}}\\
    \caption{Idea of ZSL in IoT fingerprinting} 
    \label{fig: ZSL_illus}
    \vspace{-0.7cm}
\end{figure}

Zero-shot learning (ZSL) could help to classify unseen devices. ZSL is known to work well for image classification---it leverages textual descriptions as {\it attributes} to relate unseen classes to seen classes~\cite{mishra2018generative, verma2020meta, zhao2017zero}. ZSL on images involves the use of specific attributes obtained through manual or automatic annotation. {For instance, datasets such as Caltech-UCSD~Birds-200-2011~(CUB-200)~\cite{data-CUB} and Animals with Attributes (AwA)~\cite{Data-AWA} provide pre-extracted feature representations for image descriptions.} Given the information that a giraffe is a {\it herbivore}, {\it has a long neck}, {\it has brown spots}, and {\it has ossicones on its head}, one can easily distinguish it from other animals such as pigs or cows, even without having seen a giraffe before. By leveraging the semantic relationships between different classes, the ZSL approach enables the model to recognize unseen classes by inferring their {\it attributes} in association with other classes. {The question we ask in this research is, {\em can ZSL be leveraged to carry out classification of both seen and unseen IoT devices, by mapping network traffic data to an attribute space?} Figure~\ref{fig: ZSL_illus} illustrates this concept.}

For IoT devices, we have textual descriptions, e.g., information on the product webpages or user manuals; but they do not directly translate into the domain of network traffic data. Defining attributes in the IoT domain is non-trivial, and it plays a critical role in model performance. The primary challenge in ZSL for IoT devices is the definition of suitable attributes. These attributes should possess sufficient information density to accurately represent and distinguish the traffic patterns of individual devices. In the context of network traffic analysis, one must examine the intricacies associated with packets. Network packets are characterized by a multitude of information, including IP addresses, service ports, transport protocols, inter-arrival time, etc. As data flows through the network, the accumulation of the packets results in extremely long sequences; packet sequences in hundreds or thousands or even more are common with different applications (such as browsing, email, SSH, etc.). Furthermore, the challenge posed by high dimension becomes evident when considering the diverse set of features and their potential combinations with packet sequences. This results in the creation of a high-dimensional feature space, in which each unique feature introduces an additional dimension.

While traditional statistical models and ML algorithms, such as Support Vector Machine~(SVM), Linear Regression, k-NN, and Decision Trees, have been proposed for learning network traffic behaviors (e.g., see~\cite{SLIC-2015, NADA-2018, adept}), such models encounter significant difficulties when tasked with effective processing and extracting meaningful patterns from such intricate and high dimensional data. 
Consequently, traditional machine learning algorithms, such as Support Vector Machine~(SVM), Linear Regression, and Decision Trees, encounter significant difficulties when tasked with effective processing and extracting meaningful patterns from such intricate and high dimensional data. To address these computational challenges inherent in dealing with data of long sequences and high dimensions, specialized sequence models have emerged as an effective approach. In particular, the transformer models~\cite{vaswani2017attention} have shown excellent performance in language modeling~\cite{devlin2018bert}, image classification~\cite{vit_2020}, DNA analysis~\cite{transformer_dna}, resource allocation\cite{wu2023d3t}, etc. The self-attention mechanism in transformers enables parallel evaluation with each {\em token} of the input sequence, thus eliminating the sequential dependency present in prior sequence models such as recurrent neural networks (RNNs). Transformers consider the entire context, rather than relying solely on local information, enabling a deeper understanding of context and dependencies. Building upon the encoder of the transformer architecture, we develop \atool, a self-attention mechanism designed to comprehend traffic patterns and autonomously generate concise attributes for IoT devices
(Section~\ref{Sec:attribute}).

In this work, we propose {\bf \sol}, a \underline{\bf ze}ro-\underline{\bf s}hot learning framework based on the self-attention mechanism for Io\underline{\bf T} fingerprinting. \sol involves i)~training a \underline{\bf s}elf-\underline{\bf a}ttention based \underline{\bf n}etwork feature \underline{\bf e}xtractor, i.e., \textbf{\atool}, to extract features and {\it attribute vectors} of devices, subsequently ii)~training a generative model to map attribute vectors to traffic data and generate pseudo data for unseen devices, and finally iii)~training a supervised classifier with the generated data (as illustrated in Figure~\ref{fig:zsl_arc}). The main contributions of this work are:
\begin{enumerate}
    \item We introduce a ZSL framework for IoT fingerprinting, \sol. To the best of our knowledge, it is the first generative ZSL framework for IoT fingerprinting. Our work here is also the first to leverage transformer model for learning network traffic characteristics. Based on the self-attention mechanism, \sol achieves state-of-the-art performance when compared with other semi-supervised and unsupervised learning methods. 

    \item 
    As the use of attention mechanism for network traffic understanding is new, we study its effects in classifying IoT devices and compare it with existing solutions that use LSTMs for the same purpose. We find that an attention-based mechanism improves the performance of even existing baseline models, making it a better choice for IoT fingerprinting. 

    \item We propose a new approach for generating attribute vectors for IoT devices. Unlike image classification tasks, network traffic data of IoT devices do not come with apparent class descriptions. Hence, we leverage pre-trained models to extract attribute vectors of unseen devices, providing a viable and attractive technique to overcome this challenge. We conduct experiments with varying attribute vector dimensions and identify the most suitable dimension for optimal performance. 

    \item We make the source code implementations of the models openly available to facilitate research\footnote{Code is available at: \texttt{\url{https://github.com/Binghui99/ZEST}}.}. This includes the implementation of the ZEST framework, LSTM-based baselines, as well as the data processing methods.
    
\end{enumerate}

In the following, we first present the related literature for IoT fingerprinting and the background of ZSL. In Section~\ref{sec:model architecture}, we present our ZSL framework for identifying seen and unseen devices. Performance evaluations are carried out in Section~\ref{sec:performance}.

\section{{Related Works}}\label{sec : related works}

\subsection{IoT fingerprinting}\label{sec : related IoT}

With the rapid growth of the IoT ecosystem, there has been an increasing interest in characterizing and fingerprinting (i.e., classifying) IoT devices. In recent years, several works have proposed IoT traffic analysis methods, and use supervised learning approaches to perform device classification~\cite{IoT_traffic_Apthorpe, DBLP_Markus, DBLP_Sivanathan, LSTM-device-fingerprinting-2020}. An interesting work~\cite{LSTM-device-fingerprinting-2020} is the application of a sequence model, specifically Bi-LSTM (bi-directional long short-term memory), for modeling traffic of IoT devices. This deep learning model shows a good ability to learn sequence information and achieves high accuracy in IoT device classification.

However, a supervised approach is limited in practice, since we have numerous new unseen devices entering the market regularly. Therefore, researchers proposed unsupervised methods~\cite{unsuper_2022_meditcom,PCA_kmeans_2019, ICC_unsuper} for IoT fingerprinting. Sivanathan et al.~\cite{PCA_kmeans_2019} extract key features from flow-level network traffic and use PCA (principal component analysis) to project data into lower dimensional space. As a complementary approach, the authors in~\cite{ICC_unsuper} train a VAE (variational autoencoder) with an encoder and a decoder in an unsupervised way,  subsequently leverage the encoder to compress raw data, and finally use k-means for clustering. However, unsupervised learning methods do not use all available information, such as labels of {\it seen} devices and their semantic descriptions, thus achieving only modest results (see  Section~\ref{results}).

Semi-supervised methods~\cite{DEFT-2019, CNN_semi_2020,  seeded-kmeans,semi_detection_2023}, on the other hand, utilize the available information, and they can also deal with unseen classes. Authors in~\cite{CNN_semi_2020} propose a semi-supervised method based on a CNN model (convolutional neural network) and multi-task learning. Given a few labeled data, they train a CNN to transform raw features into dense high-level features, and thus achieve a dimension reduction. An alternative semi-supervised approach called DEFT~\cite{DEFT-2019}, extracts traffic features and utilizes seeded k-means~\cite{seeded-kmeans} to conduct unsupervised clustering. Based on the clustered data, DEFT trains a supervised random forest to perform the final classification, using the cluster numbers as labels. 

The semi-supervised solutions are often based on clustering~\cite{DEFT-2019, CNN_semi_2020,  seeded-kmeans}. However, clustering has limitations, including the requirement to inform the number of clusters in advance, sensitivity to initial centroids and outliers, and unsuitability for high-dimensional and non-linear data~\cite{xu2015comprehensive, seeded-kmeans, he2022improved, yuan2019research}. In order to overcome these shortcomings, we choose to avoid clustering for our solution. Instead, our \sol framework uses a generative model for generating unseen data, based on high-level attribute definitions. Then we use the generated data to train a supervised model. To deal with high-dimensional and non-linear data, we use a deep sequence model and extract small-dimensional latent features.

\subsection{Zero-shot learning (ZSL)}\label{sec: overview ZSL}

Broadly, there are two approaches for ZSL: embedding-based methods~\cite{zhao2017zero, verma2020meta} and generative-based methods~\cite{mishra2018generative,xian2018feature, gao2020zero}. Embedding-based methods learn a high-dimensional embedding space that maps the low-level features of seen classes to their corresponding semantic vectors. This approach recognizes new classes by comparing prototypes and predicted representations of data samples in the embedding space. On the other hand, generative-based methods use samples of seen classes and semantic representations of both seen and unseen classes to generate pseudo data for the unseen classes, thus converting a ZSL problem into a supervised learning problem. 


In this work, we focus on the generative method. The authors in~\cite{mishra2018generative} propose a CVAE-based approach using a generative model to learn the probability distribution of the input space conditioned on the attribute representation of the classes. The generative model is then used to generate samples for the unseen classes based on their attribute vectors. Gao et al.~\cite{gao2020zero} propose Zero-VAE-GAN, which combines VAE and GAN to generate features for novel classes. This approach uses a dual encoder-decoder structure to map data samples into a joint feature space, improving the quality of generated samples. Our proposal \sol is inspired by~\cite{mishra2018generative}; but there are some significant differences. Firstly, the proposal in~\cite{mishra2018generative} is for the image domain; therefore, the training data utilized is accompanied by well-defined semantic information. The authors apply \texttt{word2vec} to text descriptions (e.g., from Wikipedia) for different image classes. However, such a method is unsuitable for our problem, since the text description from device manuals cannot be transferred to the network traffic domain. Therefore, we propose a different approach to extract attributes (see Section~\ref{Sec:attribute}). Moreover, in the image classification domain, there are highly proficient pre-trained models trained on large datasets, like ResNet50~\cite{resnet50}. In the absence of such pre-trained models, we train our attention-based model from scratch, and employ it to extract features from network traffic. The details of our ZSL pipeline are explained in Section~\ref{arc_design}.

\subsection{Transformer}
 Transformers, introduced by Vaswani et al.~\cite{vaswani2017attention}, are used for sequence-to-sequence learning tasks. {The self-attention mechanism is good at capturing contextual relationships within a sequence, empowering models to extract rich and informative features. Furthermore, it can be efficiently parallelized, making it suitable for modern hardware accelerators like GPUs and TPUs~\cite{BigBird-2020}.} We leverage the advancements made in transformers to enhance the performance of IoT fingerprinting. The visual transformer (ViT) proposed in~\cite{vit_2020} is a BERT-like~\cite{devlin2018bert} model for image classification that achieves superior performance on multiple benchmarks with fewer parameters than competing models, by efficiently modeling long-range dependencies between image patches and global image information. Our attention-based model is inspired by ViT. 

However, unlike the input images to ViT, network traffic comes as a data sequence. For our data pre-processing pipeline, we split network traffic into packet sequences of pre-defined length, where packets are represented by a small number of {\it raw} features (Section~\ref{Sec:traffic_representation}). For the final classification, we use an average pooling layer to get the whole sequence information. Besides, we define a special token to summarize the sequence-level network traffic information. {Our design is driven by the need to understand network traffic at both the sequence level and packet level.}

\section{\sol: model architecture}\label{sec:model architecture}

\subsection{System definition} \label{Sec:set_up_settings}

We consider a network connecting a set, $\Gamma$, of IoT devices (such as smart cameras, hubs, alarms, etc.). We use $\mathcal{S}$ to denote the set of devices already seen, and $\mathcal{U}$ for the set of unseen devices; both are mutually exclusive, i.e., $\mathcal{S} \cap \mathcal{U} = \emptyset$, $\mathcal{S} \cup \mathcal{U}=\Gamma$. For our classification system, a single data point $\mathbf{x}$ is defined as a sequence of network packets with length $n$, and each packet has $f$ features, i.e., $\mathbf{\mathbf{x}} \in \mathbb{R}^{n \times f}$. For each seen device $d \in \mathcal{S} $, we have a set of $m$ data points $\mathcal{X}^d = \{ \mathbf{x}_1^d,  \mathbf{x}_2^d, \cdots, \mathbf{x}_m^d \}$, and their corresponding labels $\mathcal{Y}^d = \{ y_1^d,  y_2^d, \cdots, y_m^d\}$. However, for $d \in \mathcal{U}$, we have data points $\mathcal{X}^d$ without the labels. {Note that, a data point representing a sequence of (say) n=200 packets, each with f=8 features, results in a feature space of 1,600---a very high dimension.}

\subsection{Traffic representation}\label{Sec:traffic_representation}


From the traffic of an IoT device, we extract, what can be referred to as, {\it raw features}, for each packet in the traffic.  Flow-level features capture the packet statistics but in a lossy way. In comparison, per-packet features provide the finest granularity of information in traffic. The raw packet features that are beneficial for device classification are: packet size, time since the last packet, the direction of the packet, transport protocol (TCP/UDP), the application protocol (HTTP/S, DNS, NTP, etc.), TLS version, src/dst IP address category, and src/dst port category. Intuitively, a single packet in itself might not present sufficient information for device classification, e.g., a TCP SYN or SYN+ACK is present in all TCP flows. 
Therefore, these per-packet features are extracted from non-overlapping fixed-length {\it sequences} of packets in a network trace of an IoT device and provided as input to the model, both in the training and inference phases. In our work, we limit the features to:

\begin{itemize}
  \item Source and destination IP addresses: Using raw IP addresses would overfit the model to the dataset used, besides creating a large latent space for representation. Instead, we use a binary value indicating whether the source/destination IP address is internal or external to the network of IoT devices. 
  \item Port representation: Between the source and destination ports, we assume the lower one is the service port, and represent only the service port. The other port is typically a random port, and therefore set to a constant number. 
  The idea is to minimize the influence of ephemeral ports.
  \item Transport layer protocol (e.g., UDP or TCP).
  \item The time since the previous network packet.
  \item The size of the packet.
  \item The direction of the packet (inbound/outbound)
\end{itemize}

\subsection{Attributes}\label{Sec:attribute}
\begin{figure}[h]
    \centering
    \subfloat{\includegraphics[scale=0.30]{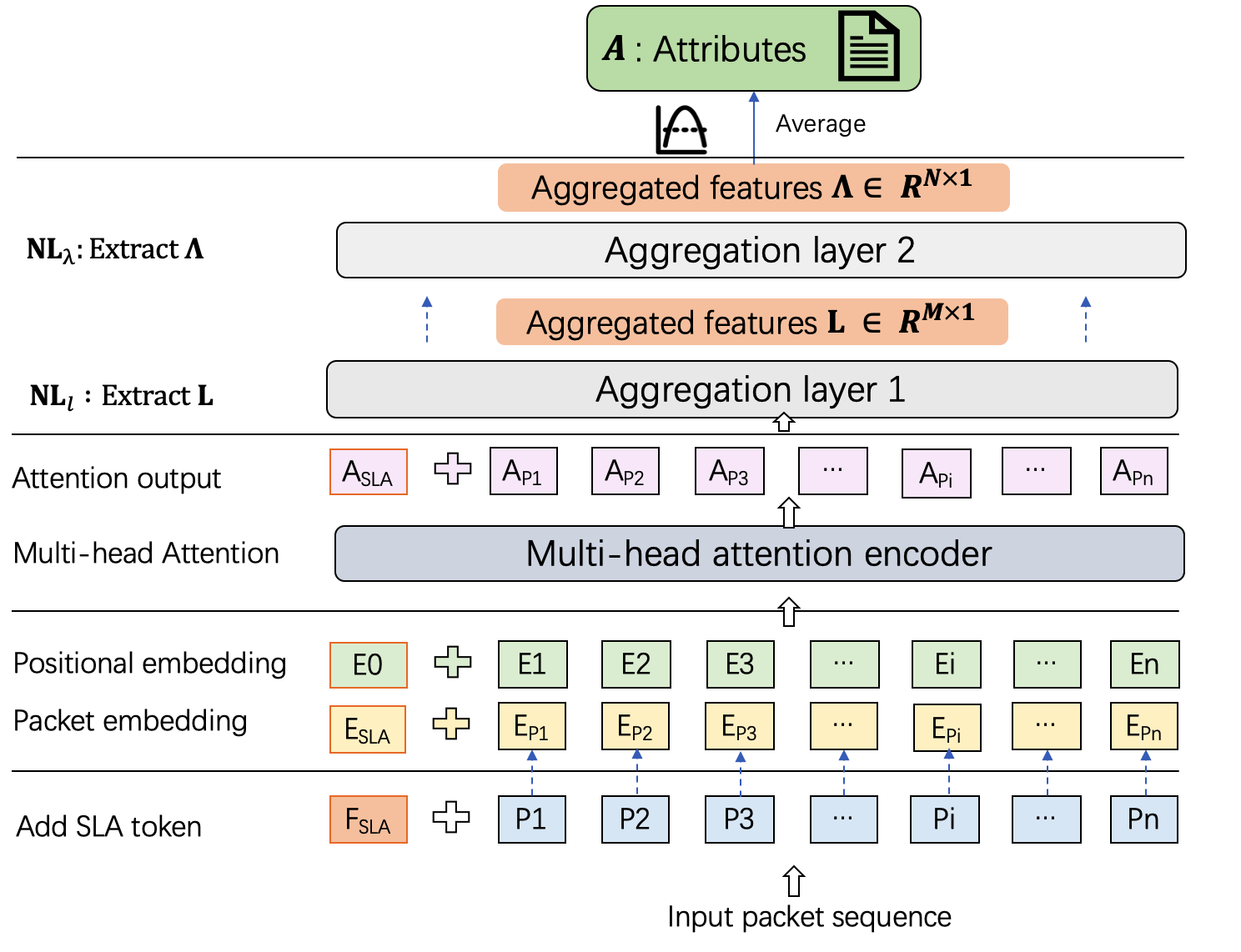}}\\
    \caption{Illustration of \atool architecture} 
    \label{fig: attr}
    \vspace{-0.5cm}
\end{figure}

To overcome the lack of meaningful textual descriptions for IoT device traffic data, we adopt a novel approach inspired by attribute-based image classification. As an analogy, attributes of a giraffe are given by the {\it wise people} who see the giraffe and describe it based on their experience of describing other animals. Specifically, we train a self-attention model on the traffic data of seen devices as ``{\it wise people}" to learn the knowledge of describing traffic patterns. 
Subsequently, when presented with traffic sequences of unseen devices, the model generates a description based on its learned knowledge, even though it has no prior knowledge of the unseen devices. The average description generated by the model is considered as the general attributes of the unseen devices. In this process, the unseen devices come with no label information. {We develop and employ a powerful self-attention model based on the encoder of transformer~\cite{devlin2018bert} to extract latent features. In Figure~\ref{fig: attr}, we present the architecture of \atool---a self-attention based feature extractor.}

The input data consists of sequences of packets with high-dimensional features, which are noisy and sparse, making it challenging for a model to recover meaningful attributes. To facilitate the learning process, we require data with high information density enabling the model to effectively map the data space to the attribute space. Therefore, we define two latent space representations to extract features at different levels for each device $d \in \Gamma$. The first one is $\mathbf{L} \in \mathbb{R}^{M \times 1}$, where $\mathbf{L}^d = \{ l_1^d,  l_2^d, \cdots, l_m^d\}$, such that, $l^d_i \in \mathbb{R}^{M \times 1}$ is the latent space representation of a single data point $\mathbf{x}_i^d$ and $m \in \mathbb{N}$ is the number of traffic sessions corresponding to device~$d$.
The second latent space is defined as $\mathbf \Lambda \in \mathbb{R}^{N \times 1} $, where $\mathbf \Lambda^d = \{ \mathbf{\lambda}_1^d, \mathbf{\lambda}_2^d, \cdots, \mathbf{\lambda}_m^d\}$, such that $\mathbf{\lambda}_i^d \in \mathbb{R}^{N \times 1}$ is another latent feature corresponding to data point $\mathbf{x}_i^d$. Based on the second latent feature, we define the attribute vector, $ \mathcal{A}^d \in  \mathbb{R}^{N \times 1}$, for both seen and unseen devices, which are embedded into a vector space representing the semantic relationship of different devices. We use the average value of $\mathbf \Lambda^d$ as the attribute vector $\mathcal{A}^d$ of corresponding device~$d$, i.e., $\mathcal{A}^d$ = $\frac{\sum_{i}^m \mathbf \lambda_i^d}{m}$.

\subsection{Architecture design}\label{arc_design}

\begin{figure}[h]
    \centering
    \subfloat[\atool training and attribute extraction]{\label{fig:Zest_part1}\includegraphics[scale=0.27]{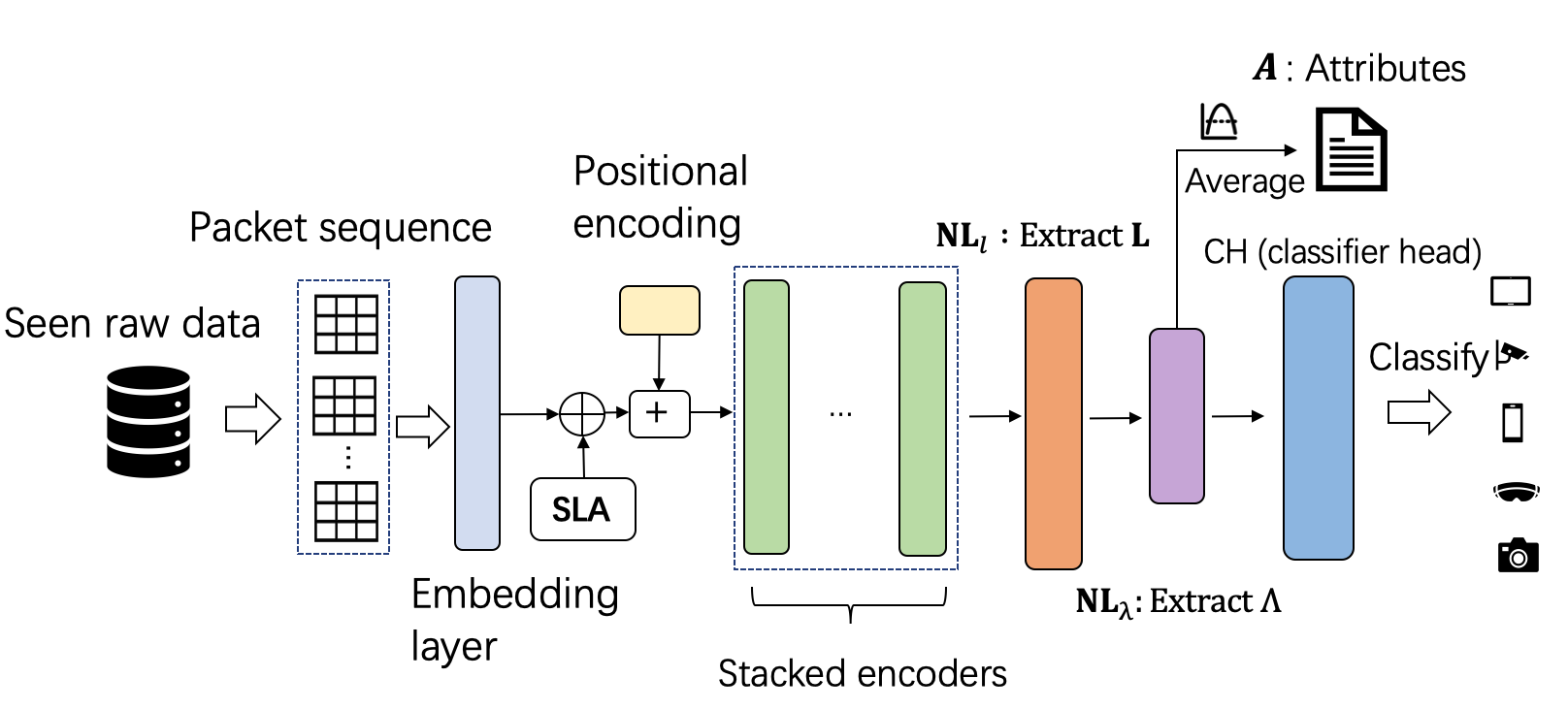}}\\
    \subfloat[Generative model training (CVAE)]{\label{fig:Zest_part3}\includegraphics[scale=0.32]{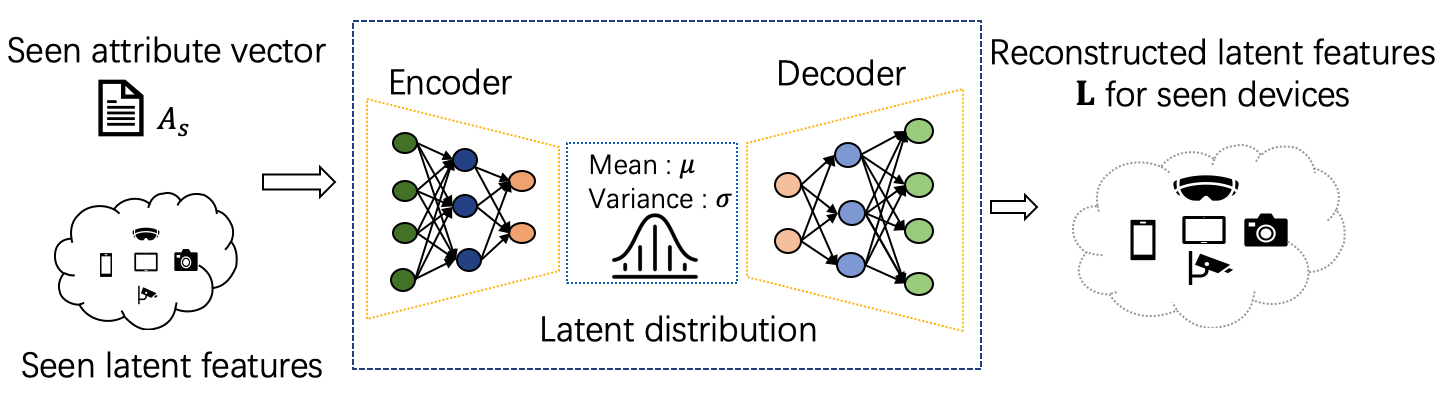}}\\
    \subfloat[Pseudo data generation (CVAE) \& supervised classifier training ]{\label{fig:Zest_part4}\includegraphics[scale=0.32]{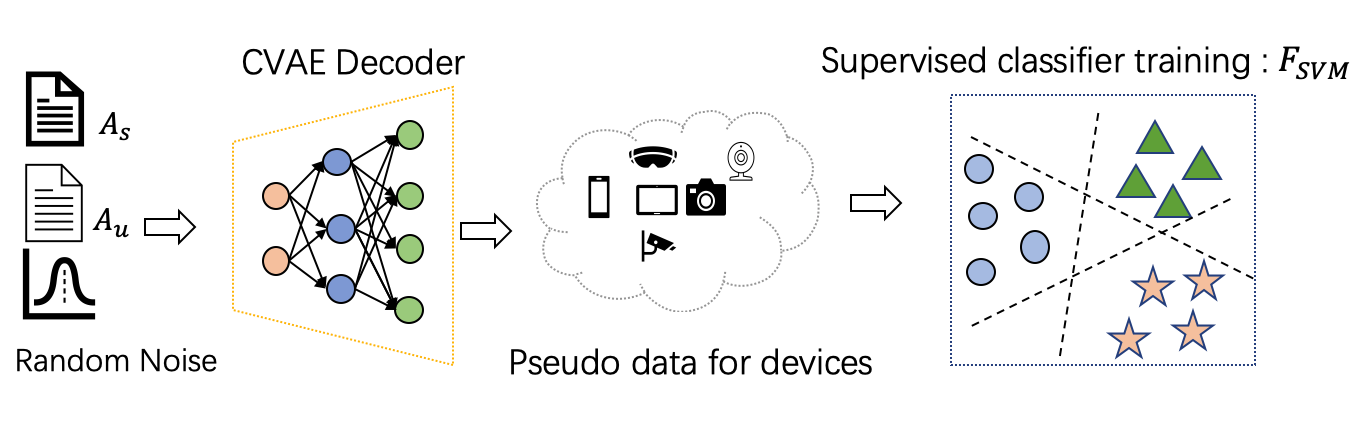}}\\
    \caption{\sol architecture} 
    \label{fig:zsl_arc}
    \vspace{-0.6cm}
\end{figure}

The architecture of \sol is depicted in Figure~\ref{fig:zsl_arc}. It consists of four phases: feature extractor training, feature and attribute extraction, generative model training, and training of a final supervised classifier.

\noindent \textbf{{\atool} training (Algorithm 1)}. {\atool} is used to transform the input data into a latent feature space, as shown in Figure~\ref{fig:Zest_part1}. The deep sequence classifier is pre-trained on a significant amount of {\it seen} device data. After comparing well-known sequence models (namely, LSTM and transformer), we find that the transformer model has better ability to extract traffic patterns (see Section~\ref{Sec:attribute}) as well as the lowest inference time. Therefore, we design \atool, based on transformer (encoder), as the feature extractor for \sol. The {\it raw} data for seen devices undergo a transformation into packet sequences with a sequence length of $n$ = 200 packets. Recall from Section~\ref{Sec:traffic_representation} that each packet is represented using $f$ = 8 features. The steps of 
{\atool} training are listed in Algorithm~\ref{alg:Transformer training}.

First, we project packet sequences using a normal linear layer to create an embedding. Then, we add the special \texttt{SLA}~(Sequence Level Aggregation) token and the positional embedding. {The learnable \texttt{SLA} token is specifically} designed for summarizing sequence-level features, facilitating the integration of both packet-level and sequence-level information. It is randomly initialized and added before the packet embedding. This integration results in powerful representations of network traffic. Additionally, we employ a positional embedding, denoted as $\mathbf{P}$, to provide the model with information about the original packet positions. This embedding is initialized as a random vector, learned by the model, and subsequently added to the packet embedding.

\begin{algorithm}
\begin{footnotesize}  
\renewcommand{\algorithmicrequire}{\textbf{Input:}}
\renewcommand{\algorithmicensure}{\textbf{Output:}}
\caption{{\atool training}}\label{alg:Transformer training}
\begin{algorithmic}[1]
\Require $\mathcal{X}^d$; $\mathcal{Y}^d$ $\forall \: d \in \mathcal{S}$ 
\Ensure  $\mathcal{\xi}_{\text{\atool}}$

\State Randomly initialize parameters in \atool

\For {$(\mathbf{b}_x, \mathbf{b}_y)$ : sample a batch from $\mathcal{X}$ and $\mathcal{Y}$} 
        \State $ \mathbf{E} =  \texttt{SLA} \oplus \text{Embedding} (\mathbf b_x) $ \Comment{Concatenate \texttt{SLA}}
        \State $\mathcal{E}  \gets \mathbf{P} + \mathbf{E} $ \Comment{Add positional embedding}
        \For{each stacked encoder $\nu$} \label{begin_of_encoder}
        \State $ R^1 \gets \text{Multi\_head\_attention} (\text{Norm}(\mathcal{E}))$ 
        \State $ R^2 \gets \text{MLP} (\text{Norm}(\mathcal{E} + R^1)) $
        \State $\mathcal{E} \gets R^2 + (\mathcal{E} + R^1) $
        \EndFor \label{end_of_encoder}
        \State $l_i = \text{$\mathbf{NL}_l$}(\text{Average\_pooling}(\mathcal{E} ))$;~~$\mathbf \lambda_i = \text{$\mathbf{NL}_{\lambda}$} ( l_i)$ 
        \State $ \text{Predictions} \gets \mathbf{Softmax} (\text{Classifier\_head}(\mathbf \lambda_i))$
        \State $ \text{Loss} \gets \mathbf{Cross\_entropy} (\text{Predictions}, \mathbf b_y) $
        \State $\theta \gets \theta + \mathbf{Compute\_update}(\text{Loss}) $
\EndFor

\State $\mathcal{\xi}_{\text{\atool}}$ = ${\xi(\theta)}$ \Comment{Trained {\atool}}

\State \Return  $\mathcal{\xi}_{\text{\atool}}$

\end{algorithmic}
\end{footnotesize} 
\end{algorithm}

As a next step in our {\atool} architecture, we employ a stack of encoders~\cite{vit_2020} to learn the mapping from the {\it raw} data space to a latent space by classifying seen devices, see line no. \ref{begin_of_encoder}-\ref{end_of_encoder} in Algorithm~\ref{alg:Transformer training}. 
After the encoder stack, we adopt an average pooling layer to aggregate the features from both packet-level and sequence-level. Next we add two different dense layers \text{$\mathbf{NL}_l$} and \text{$\mathbf{NL}_{\lambda}$} to $\mathcal{C}$ for deriving latent features $\mathbf L$ and $\mathbf \Lambda$, respectively.
Finally, we pass the latent features to a classifier head and use $\mathbf{Softmax}$ to get the prediction probabilities per device class. The final step of the forward path is the calculation of the loss. Then we perform back-propagation to compute the weight updates. This procedure is repeated for several epochs until we get our final trained model,~$\mathcal{\xi}_{\text{\atool}}$.

\begin{algorithm}[h]
\begin{footnotesize}

\renewcommand{\algorithmicrequire}{\textbf{Input:}}
\renewcommand{\algorithmicensure}{\textbf{Output:}}
\caption{\atool feature and attribute extraction}\label{alg:feature_extractor_training}
\begin{algorithmic}[1]
\Require $\mathcal{X}$;  $\mathcal{Y}$; $\mathcal{\xi}_{\text{\atool}}$

\Ensure $\mathcal{C}$: feature extraction model; $\mathcal{A}$: Attribute vector;  $\mathbf L$: latent features

\State $\mathcal{C} \gets $ Remove\_classifier\_head ($\mathcal{\xi}_{\text{\atool}}$)
\State $\mathcal{C}_l \gets \text{Remove}\_\mathbf{NL}_{\lambda} (\mathcal{C}) $   \Comment{Remove layer $\mathbf{NL}_{\lambda}$  }

\State $\mathcal{C}_{\lambda} \gets  \text{Remove}\_\mathbf{NL}_{l} (\mathcal{C}_l)$\Comment{Remove layer $\mathbf{NL}_l$ }

\For {$\mathbf x_i$ in $\mathcal{X}$, $\forall \: i = 1,2,\cdots, m$}\label{start of attribute}
        \State $l_i = \mathcal{C}_l(\mathbf x_i)$   \Comment{Extract feature $l \in \mathbb{R}^{M\times1}$}
        \State $\lambda_i = \mathcal{C}_{\lambda}(\mathbf x_i)$ \Comment{Extract feature $\mathbf{\lambda} \in \mathbb{R}^{N\times1}$}
\EndFor 

\State $\mathbf L$ = $\{ l_1, l_2, \cdots, l_m\}$
\State $\mathbf \Lambda$ = $\{\mathbf \lambda_1, \mathbf \lambda_2, \cdots, \mathbf \lambda_m\}$
\State $\mathcal{A}$ = $\frac{\sum_{i}^m \mathbf \lambda_i} {m}$ \Comment{Take average of $\mathbf \Lambda$} \label{end of attribute}

\State \Return $\mathcal{C}$, $\mathcal{A}$, $\mathbf L$
\end{algorithmic}
\end{footnotesize}
\vspace{-0.1cm}
\end{algorithm}

\noindent \textbf{Feature and attribute extraction (Algorithm 2)}.
In this stage of \sol, we feed $\mathcal{X}^d, \forall \: d \in \mathcal{U}$ to $\mathcal{\xi}_{\text{\atool}}$, an attention-based encoder trained only on seen classes (the ``{\it wise people}"), to derive the attribute vectors of the unseen devices (without labels); see Figure~\ref{fig:Zest_part1}.
To achieve this, we first remove the classification head of the trained {\atool} $\mathcal{\xi}_{\text{\atool}}$ to get the feature extraction model~$\mathcal{C}$. 
For a device $d \in \Gamma$, we remove the layer \text{$\mathbf{NL}_\lambda$} from $\mathcal{C}$ to get $\mathcal{C}_l$ for $\mathbf{L}^d$ extraction. 
Then we remove layer \text{$\mathbf{NL}_l$} from $\mathcal{C}_l$ to get $\mathcal{C}_\lambda$ for $\mathbf{\Lambda}^d$. The attribute extractor takes the average value of $\mathbf{\Lambda}^d$ to compute attribute vector $\mathcal{A}^d$ for device $d$. Line no.~\ref{start of attribute}-\ref{end of attribute} in Algorithm~\ref{alg:feature_extractor_training} illustrate the process of extracting attribute vectors for both seen and unseen devices.

\begin{algorithm}[h]
\begin{footnotesize}
\renewcommand{\algorithmicrequire}{\textbf{Input:}}
\renewcommand{\algorithmicensure}{\textbf{Output:}}
\caption{CVAE and SVM training}\label{alg:CVAE model training}
\begin{algorithmic}[1]
\Require $\mathcal{A}^s$; $\mathbf L^s, \forall \: s \in \mathcal{S}$;  $\mathcal{A}^u, \forall \: u \in \mathcal{U}$; 
\Ensure  $\mathcal{F}_{\text{SVM}}$ : Trained SVM classifier

\State Randomly initialize $\text{Encoder}_{\text{cvae}} $, $\text{Decoder}_{\text{cvae}}$ \label{CVAE_begin}

\For {$\mathbf b$: sample a batch from $\mathbf L^s$} 
        \State $\mu , \sigma  \gets \text{Encoder}_{\text{cvae}}$ ($\mathbf{b},\mathcal{A}^s) $ 
        \State Sample $\mathbf{z}$ form $\mathcal{N} \sim (\mu, \sigma)$ 
        \State $\mathbf{\hat{b}} \gets \text{Decoder}_{\text{cvae}}$ ($\mathbf{z}, \mathcal{A}^s)  $ \Comment{Reconstruction}
        \State Loss $\gets$ $|(\mathbf{\hat{b}} - \mathbf{b})|$ + KL($\mathcal{N}(\mu, \sigma),\mathcal{N} (0,1)$) 
        \State $\theta \gets \theta + \mathbf{Compute\_update}(\text{Loss}) $
\EndFor

\State $\mathcal{D}$ = $\text{Decoder}_{\text{cvae}}(\theta)$ \Comment{Trained decoder} \label{CVAE_end}
\State Sample random noise $\tau$ form  $N\sim(0, 1)$  
\State $\mathcal{P}_u \gets \mathcal{D} ( \tau, \mathcal{A}^u )$; $\mathcal{P}_s \gets \mathcal{D} ( \tau, \mathcal{A}^s )$ \Comment{Pseudo data generation }
\State $\mathcal{P} = \mathcal{P}_u \cup \mathcal{P}_s$ 

\State $\mathcal{F}_{\text{SVM}} \gets \text{SVM\_fit} (\mathcal{P})$ \Comment{SVM training}

\State \Return  $\mathcal{F}_{\text{SVM}}$

\end{algorithmic}
\end{footnotesize}
\end{algorithm}

\noindent \textbf{Generative model training (Algorithm 3)}. The goal of this stage is to generate pseudo data for unseen IoT devices. For this purpose, we leverage a well-known generative model, namely a CVAE (conditional variational autoencoder). The CVAE contains two parts: an encoder and a decoder. The encoder compresses the latent features conditioned on the attributes of devices. Next, the decoder reconstructs the latent features, based on IoT device attributes. CVAE training is based only on seen devices' latent features $\mathbf{L}^s,  \forall \: s \in \mathcal{S}$ and attribute vector $\mathcal{A}^s, \forall \: s \in \mathcal{S}$, see line no.~\ref{CVAE_begin}-\ref{CVAE_end} in Algorithm~\ref{alg:CVAE model training}. For training of the CVAE architecture, we use two losses functions, namely the reconstruction loss and the KL divergence \cite{mishra2018generative} between the latent distribution and normal Gaussian distribution. 
During the training phase, the decoder $\mathcal{D}$ of the CVAE learns to map attribute vectors to the initial latent features. We use this ability of $\mathcal{D}$ to generate pseudo data $\mathcal{P}_u$ of unseen device classes based on attributes $\mathcal{A}^u, \forall \: u \in \mathcal{U}$, by querying random noise $\tau$ sampled from the normal Gaussian distribution. This unseen pseudo data is then used in the next stage for a supervised classifier. 
During our experiments, we found that the final classifier faces a bias towards seen classes when we use real seen device data, an observation that has already been made in \cite{mishra2018generative}. Therefore, we generate an equal number of pseudo data $\mathcal{P} = \mathcal{P}_u \cup \mathcal{P}_s$ for devices from both seen and unseen classes. These steps are illustrated in Figure~\ref{fig:Zest_part3}.

\noindent \textbf{Supervised classifier training}. After generating labeled data for both seen and unseen devices, we employ a final supervised classifier, support vector machine (SVM), to predict the IoT device label; see Figure~\ref{fig:Zest_part4}. Finally, we test the performance of the trained SVM model in classifying real traffic data from both seen and unseen devices.

\section{Performance evaluation} \label{sec:performance}

In this section, we evaluate \sol and compare it with existing semi-supervised learning approaches. Typically, the performance evaluation happens in two different settings: one is ZSL, and another one is GZSL (generalized zero-shot learning). In both settings, the training set contains only the labeled data of seen classes, $\{\mathcal{X}^d, \mathcal{Y}^d, \forall \: d \in \mathcal{S}\}$. The objective of ZSL is to classify the unseen devices~$\mathcal{U}$, which are not seen during the training stage. Different from ZSL, the test data of GZSL come from both seen~$\mathcal{S}$ and unseen devices $\mathcal U$. To sum up, the goals for ZSL and GZSL are to learn the classifications: $f_{\text{ZSL}}: \mathcal{X} \rightarrow \mathcal{U}$ and $ f_\text{GZSL}: \mathcal{X} \to \mathcal{U} \cup \mathcal{S}$, respectively.

\subsection{Dataset for evaluation}

We carry out evaluations on the publicly available UNSW 2018 dataset~\cite{sivanathan2018classifying}. The appeal of this dataset stems from the fact that it is relatively large (27 GB of data) and contains a good number of IoT devices (28 device types). The UNSW dataset is composed of raw \texttt{pcap} data collected during a period of 61 days. We ignore the gateway device, as it only acts as an intermediary between IoT devices and the Internet. Due to the imbalance in data per device type, for both ZSL and GZSL, we use only the 12 device classes with the most number of data points; of these, 10 random devices form the {\it seen} category, and the rest form the {\it unseen} category. The dataset contains more than a million data points across 12 devices, with the minimum being 14,034 data point for a device, and the maximum being 197,876 data points. We do not include the {\it unseen} devices in the first supervised training step. The seen and unseen devices are selected randomly five times, and we report the average result over all five experimental runs. 



\subsection{Baselines for comparisons}

We compare \sol with the baselines on the benchmark dataset. 
All the training and testing of baselines and \sol are performed on an Nvidia Tesla T4 GPU. 
As Bi-LSTM has been shown to have excellent performance in supervised IoT fingerprinting~\cite{LSTM-device-fingerprinting-2020}, we use Bi-LSTM as the feature extractor for all the baselines. 
We elaborate: sequences of packets are transformed to a smaller dimension feature vector $\mathbf{L} \in \mathbb{R}^{20 \times 1}$ using a Bi-LSTM model. These feature vectors produced by the Bi-LSTM model are then provided as input to the respective models of the baselines. Therefore, just like \sol, all baselines take the transformed and reduced feature vectors as input, thus making sure the comparisons are fair and are at the algorithmic level. 
Note, the Bi-LSTM model, as well as the {\atool} model for \sol, are trained on the seen classes. We now describe the baselines:

\begin{enumerate}

    \item \textbf{VAE-K}: This is originally proposed as an unsupervised learning method in~\cite{ICC_unsuper}, employing VAE to extract features and applying k-means to perform clustering subsequently. In our implementation, the VAE encoder compresses the features $\mathbf{L}$ generated by the Bi-LSTM model to the same dimension as $\mathbf{\Lambda}$, another latent feature with a lower dimension. It then employs k-means to cluster data into 12 groups, the same as the total number of seen and unseen devices. Since the Bi-LSTM model is trained on seen classes, the model is now semi-supervised, and therefore also more capable than the unsupervised counterpart.

    \item \textbf{SeqCR}: This is a semi-supervised sequence clustering model. Rather than using the feature vectors $\mathbf{L}$ generated by Bi-LSTM directly, it adds one more layer to extract ${\mathbf{\Lambda}}$ from the initial packet sequences. Thus, avoiding VAE, we still obtain feature vectors of a much lower dimension, $\mathbf{\Lambda}$. Subsequently, we apply k-means clustering to assign the latent features ${\mathbf{\Lambda}}$ to 12 groups. The cluster centers are randomly initialized.

    \item \textbf{SeqCS}: This is an extension of {\it SeqCR} and a semi-supervised sequence-based clustering model. Unlike {\it SeqCR}, it initializes the k-means cluster centers using the attribute vector $\mathcal{A}$, which is the average value of $\mathbf{\Lambda}$ (from both seen and unseen devices), and uses seeded k-means~\cite{seeded-kmeans} for clustering. This helps improve the classification performance since the attribute vectors contain auxiliary information for each cluster. 

    \item \textbf{DEFT}: This is a semi-supervised learning method proposed in \cite{DEFT-2019}, and can be seen as an extension of {\it SeqCS}. After getting initial clustering results based on {\it SeqCS}, DEFT trains another random forest for classification based on the clustered dataset with $\mathbf{\Lambda}$, as well as their cluster labels. However, if the initial clustering done by {\it SeqCS} is of poor quality, it can negatively impact the final performance.
\end{enumerate}

We make the code of \sol and all the baselines public~\cite{zest-git}.

\subsection{Results}\label{results}

To the best of our knowledge, our work is the first to apply the generative ZSL framework on network traffic. Therefore, besides studying the efficacy of \sol for IoT device classification, we conduct various experiments: we {carry out ablation experiments to }investigate the efficiency of the {\atool} model, we perform {\atool} hyper-parameter tuning, and we analyze the ability of \sol to handle varying numbers of unseen devices.

\subsubsection{Comparison of baselines vs. \sol}

\begin{figure}[h]
\centering
\captionsetup{justification=centering}
\includegraphics[width=0.36\textwidth]{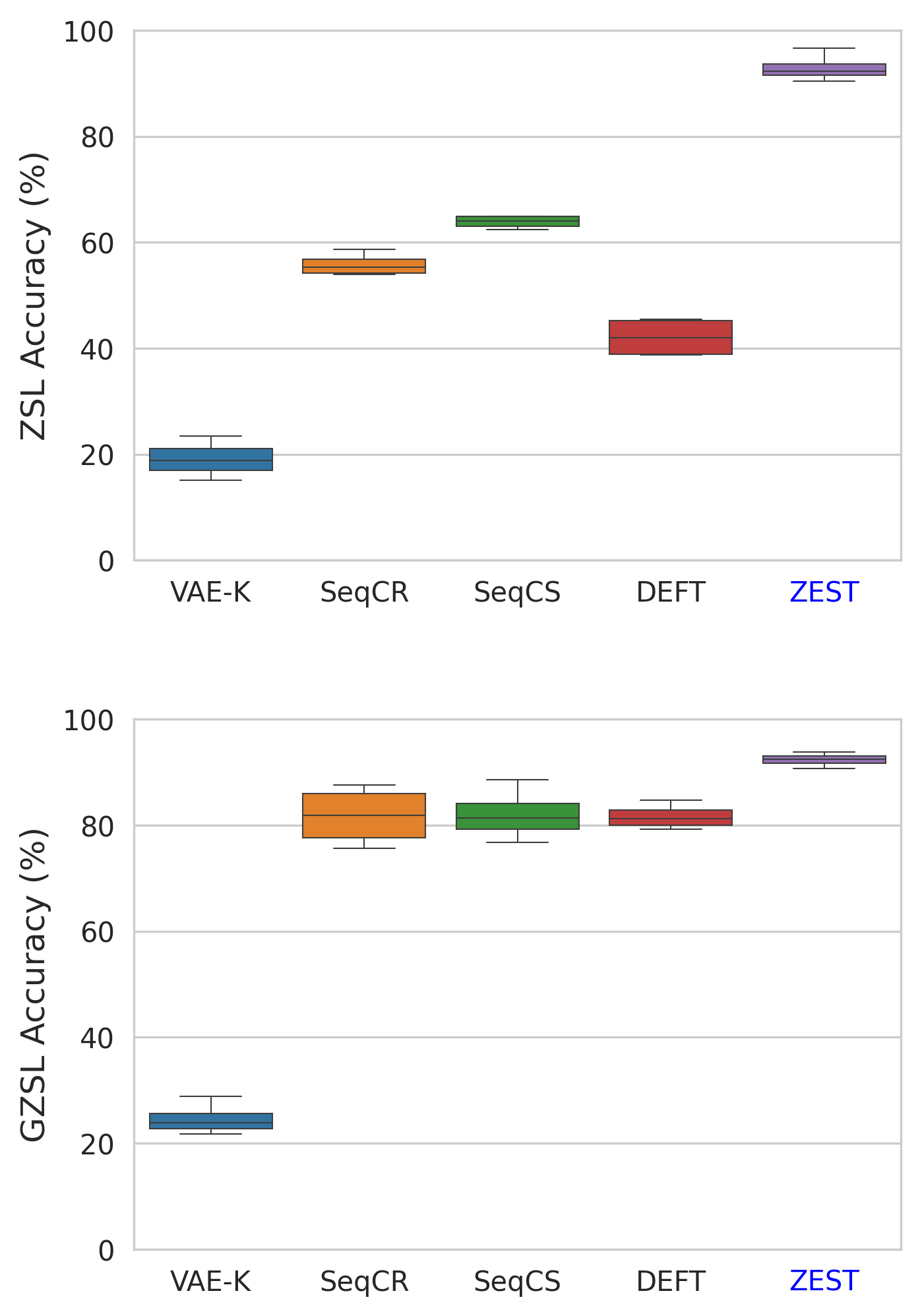}
\caption{Comparison of baselines and \sol}
\label{fig: box_all_baselines}

\end{figure}

We evaluate the performance using two evaluation metrics: ZSL accuracy and GZSL accuracy. We consistently use a 5:1 ratio of seen to unseen devices (10 {\it seen}, 2 {\it unseen}) in our experiments. As depicted in Figure~\ref{fig: box_all_baselines}, in the ZSL setting, SeqCS achieves the best accuracy of $\sim 64\%$ among the baselines. \sol achieves superior accuracy of $\sim 93\%$, meaning we get nearly
\textbf{30\%} absolute improvement with \sol over the best performing baseline. As for GZSL, the accuracy of \sol is $\sim 92\%$, which is an absolute \textbf{10\%} improvement when compared to the next best-performing baseline DEFT.

\subsubsection{Analysis of {\atool} architecture}
We investigate the impact of the number of stacked encoders, denoted as $e$, and the number of attention heads, denoted as $h$, in our {\atool} model. Using a stack of encoders helps the {\atool} model to learn more complex relationships, similar to a CNN with many convolution layers. Within one encoder, the attention mechanism is applied multiple times in parallel, giving us a so-called multi-head attention layer. This mechanism enables the model to jointly consider different patterns to pay attention to, while also offering great possibility for parallelization during training~\cite{vaswani2017attention}.

Initially, we set the number of attention heads {in \atool} to $8$ and vary the encoder stack size for $e = \{1,2,4,6,8\}$. The batch size is set to 64 and we train 20 epochs with a learning rate of 0.0005. Our analysis considers the resource consumption, which takes into account the parameter size and training time, as well as the supervised training accuracy of the different settings. The results presented in Table~\ref{tab:num_encoder} demonstrate that increasing the number of encoders leads to a proportional increase in training time and parameter size. However, the prediction accuracy remains around $98.8\%$ when the number of encoders ranges from $2$~to~$8$. Hence, after considering the trade-off between resources and accuracy, we set $e=2$.

\begin{table}[ht]
    \centering
    \caption{Comparison of different encoder stack sizes}
    \resizebox{0.9\columnwidth}{!}{%
    \begin{tabular}{cccc}
        \hline
        Number of  & Parameter & Training time & Accuracy  \\
        encoders &  size (MB) & /epoch (min) & (\%)\\
        \hline
        1 & 2.81 & 2.10 & 98.23 \\
       
        2 & 5.56 & \textbf{3.90} & \textbf{98.84} \\
        
        4 & 11.09 & 7.85 & 98.76 \\
        
        6 & 16.61 & 11.10 & 98.81 \\
        
        8 & 22.14 & \textbf{14.90} & \textbf{98.86} \\
        \hline
    \end{tabular}
    }
    \label{tab:num_encoder}
    \vspace{-0.2cm}
\end{table}

Having fixed the encoder stack size, we now vary the number of attention heads to $h=\{1,2,4,8\}$. The results presented in Table~\ref{tab:num_heads} show that increasing the number of attention heads leads to higher accuracy. This is because more attention heads enable the model to capture a broader range of aspects in the traffic pattern. Therefore, we choose $h=8$. In conclusion of our hyperparameter analysis, we find that the optimal encoder stack size is relatively small at $e=2$. On the other hand, the multi-head attention layers should be relatively large, each containing $h=8$ parallel attention heads.

\begin{table}[ht]
    \centering
    \captionsetup{justification=centering}
    \caption{{Comparison of different number of attention heads in \atool}}
    \resizebox{0.95\columnwidth}{!}{%
    \begin{tabular}{cccc}
        \hline
        Number of  & Parameter & Training time & Accuracy  \\
        attention heads &  size (MB) & /epoch (min) & (\%)\\
        \hline
        1 & 1.24 & 1.42 & 97.81 \\
        
        2 & 1.86 & \textbf{1.68} & \textbf{98.36} \\
        
        4 & 3.09 & \textbf{2.37} & \textbf{98.45} \\
        
        8 & 5.56 & \textbf{3.90} & \textbf{98.84} \\
        \hline
    \end{tabular}
    }
    \label{tab:num_heads}
\end{table}

\subsubsection{Comparison of {\atool} vs. LSTM} \label{Sec:Trans}

In this study, we first investigate the impact of using {\atool} in comparison to a Bi-LSTM for feature extraction in supervised learning. We consider all 28 classes and create a train/val/test data split in the ratio 60:20:20, via random shuffling. Table~\ref{tab: LSTM_Transfomer comparison} presents the prediction accuracy and inference time of these two models. From the results, we observe that the {\atool} model achieves a slightly higher accuracy, yet at a much lower prediction time than the Bi-LSTM model. To further study the effectiveness of using {\atool}, we conduct another experiment on the baseline models, where we replace the underlying Bi-LSTM sequence models with the {\atool} architecture.

\begin{table}[ht]
    \centering
    \captionsetup{justification=centering}
    \caption{Comparison of {\atool} and Bi-LSTM}
    \resizebox{0.9\columnwidth}{!}{%
    \begin{tabular}{ccc}
        \hline
	{Method} &	{Bi-LSTM} & {{\atool}} \\
        \hline
        {Classification Accuracy (\%)} & 97.46 & \textbf{98.84}  \\
        
        {Inference Time (ms)} & 0.59 & \textbf{0.12}  \\
        \hline
    \end{tabular}
    }
    \label{tab: LSTM_Transfomer comparison}
    \vspace{-0.25cm}
\end{table}

\begin{figure}[h]
    \centering
    \captionsetup{justification=centering}
    \subfloat{\label{fig2}\includegraphics[scale=0.5]{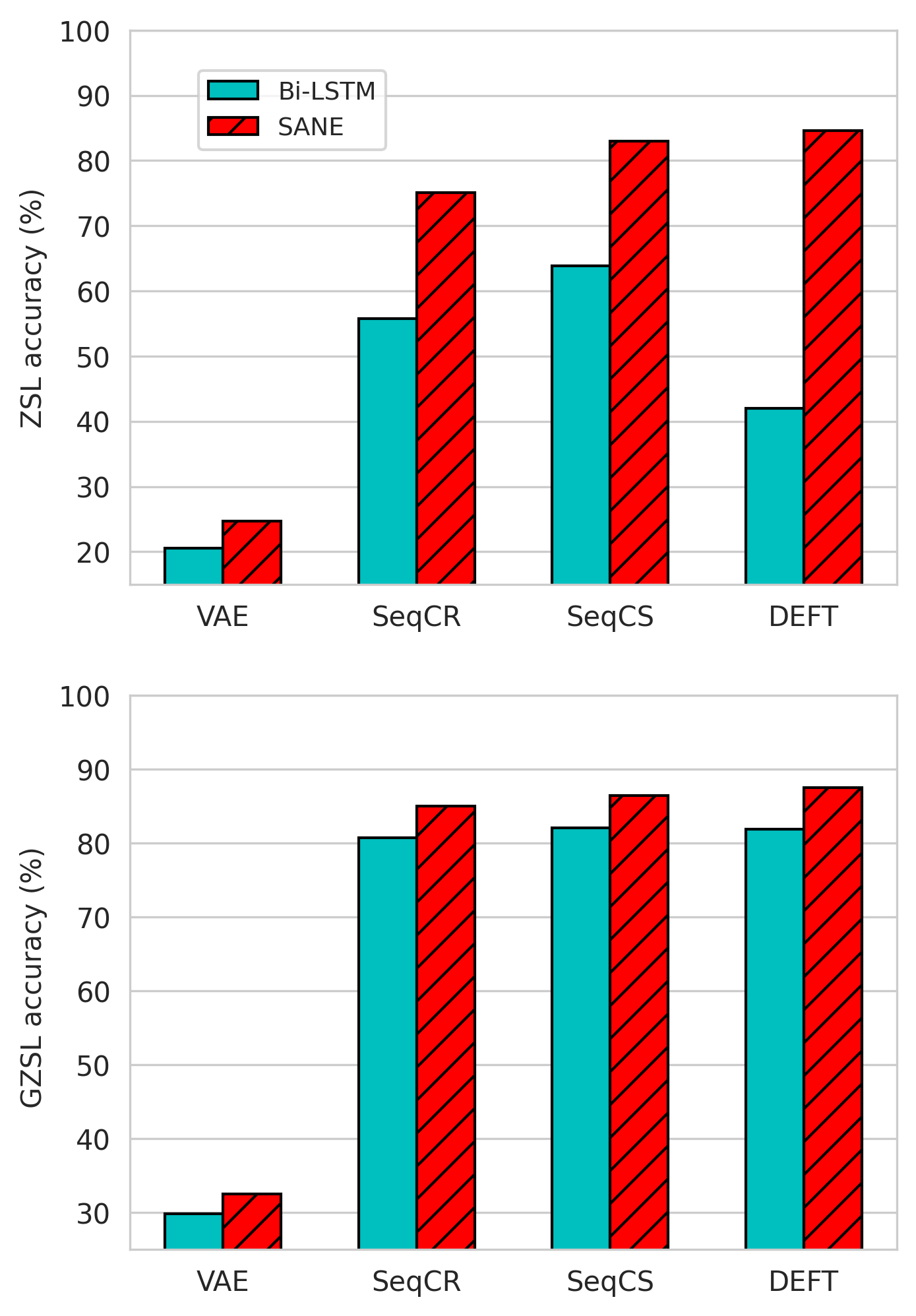}}\\
    \caption{{\atool} vs. Bi-LSTM for baselines} 
    \label{fig:comparsion_between_lstmandtf}
    \vspace{-0.2cm}
\end{figure}

Figure~\ref{fig:comparsion_between_lstmandtf} illustrates the ZSL and GZSL accuracy when feature extraction is done by Bi-LSTM and {\atool}. In the ZSL setting, {\atool}-based models yield an average absolute improvement of approximately \textbf{20\%} over various baselines, with a maximum improvement of about \textbf{40\%} observed for DEFT. In the GZSL setting, {\atool}-based models achieve an average absolute accuracy gain of about \textbf{5\%}, compared to Bi-LSTM. These results suggest that {\atool} is capable of extracting more informative features, leading us to select it as our preferred feature extraction method in \sol. Additionally, as is evident from Figure~\ref{fig: box_all_baselines} and Figure~\ref{fig:comparsion_between_lstmandtf}, \sol also outperforms {\atool}-based baselines. \sol brings an average absolute improvement of about 10\% and 5\% for ZSL and GZSL settings, respectively, when compared with the best {\atool}-based baseline, DEFT.

\subsubsection{Comparison of different attribute dimensions}

\begin{figure}[h]
    \centering
   \captionsetup{justification=centering}
    \subfloat[Comparison of different attribute dimensions]{\label{fig:attribute dimension}\includegraphics[scale=0.6]{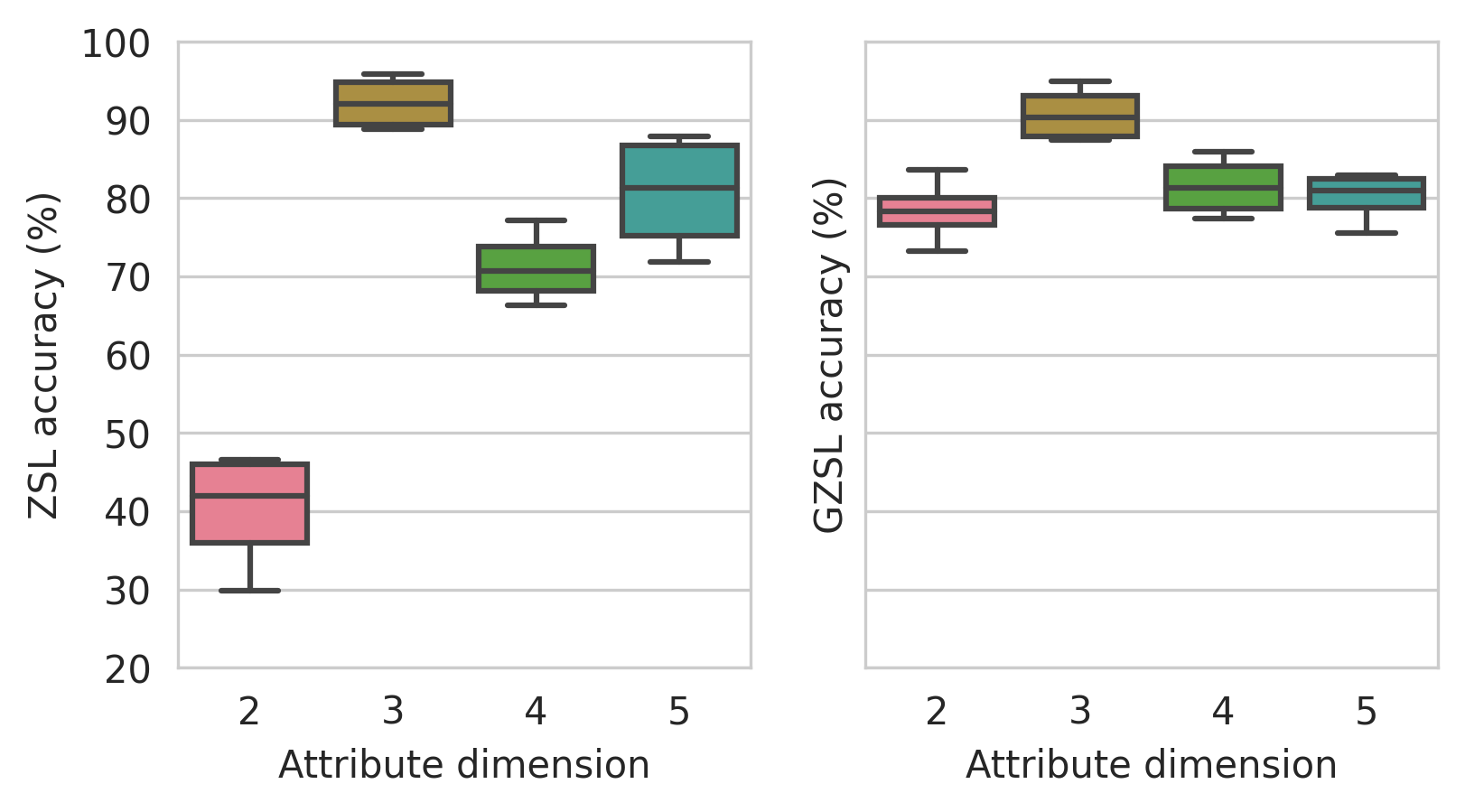}}\\
     \subfloat[Comparison of varying number of unseen classes]{\label{fig:unseen}\includegraphics[scale=0.6]{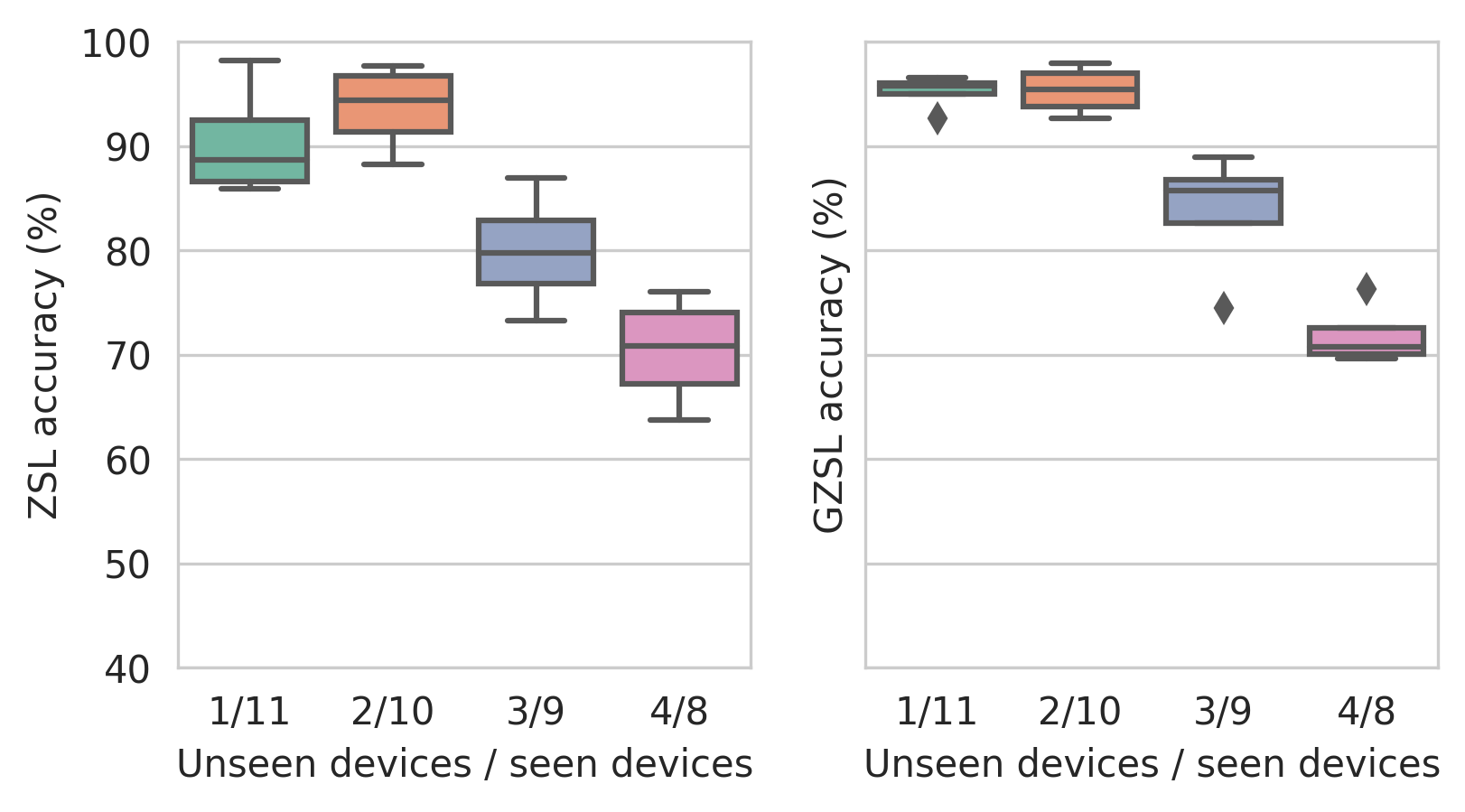}}\\
    \caption{Analysis for attribute vector dimension and number of unseen classes} 
    \vspace{-0.4cm}
\end{figure}

Here we search for the optimal attribute dimension for \sol. To assess the quality of the reconstructed data from the CVAE, we experiment by varying the dimension of the attribute vector $\mathcal{A}$, specifically, we use $|\mathcal{A}| = \{2,3,4,5\}$. As shown in Figure~\ref{fig:attribute dimension}, we observe that the best performance is achieved when the dimension is 3 in both ZSL and GZSL settings. This trend of increasing accuracy followed by decreasing accuracy is reasonable---while a higher-dimensional attribute vector represents more traffic information, it can also make the mapping between attributes and traffic data more challenging. 

\subsubsection{Varying the number of unseen classes}
We investigate \sol's ability to handle multiple unseen classes. In Figure~\ref{fig:unseen}, we compare its performance when varying the number of unseen devices from 1 to 4, for a total of 12 classes. As the number of unseen classes increases, both ZSL and GZSL accuracy values tend to decrease. When the ratio changes from $2/10$ to $3/9$, there is a sharp decrease for both ZSL and GZSL. Lower ratios of unseen classes to seen classes facilitate a more effective mapping of attribute vectors to traffic data. In other words, \sol requires a significant number of devices in the seen category to classify unseen devices. A rigorous validation of this observation using 50-100 devices is worth a different study. In practice, however, given that MAC addresses of devices remain static over a duration, the model can effectively classify traffic from a small number of unseen IoT devices.

\vspace{-0.2cm}
\section{Conclusions}\label{sec: conclusion}

In this work, we develop a novel zero-shot learning framework (ZSL), called \sol, for IoT fingerprinting. As the first attempt to work on ZSL for network traffic modeling, we analyze the effectiveness of the self-attention based model for extracting traffic features, in this framework. Our experiments show that \atool yields higher accuracy, lower inference time, and better feature extraction ability in comparison to Bi-LSTM. Since ZSL relies on class-specific attributes, which are typically not present for traffic-based IoT device classification, we propose a novel attention-based approach, i.e., \atool, to automatically compute attributes for any IoT device. Finally, we compare our \sol framework with four baselines from both the unsupervised and semi-supervised domains. Our results demonstrate that \sol achieves state-of-the-art performance, with an absolute accuracy improvement of about 30\% and 10\% for ZSL and GZSL, respectively.

\bibliographystyle{IEEEtran}
\bibliography{ref}
\end{document}